\documentclass[twocolumn,showpacs,preprintnumbers,amsmath,amssymb,nofootinbib,pre,twocolumn]{revtex4}
\usepackage{graphicx}
\usepackage{dcolumn}
\usepackage{amsmath}
\usepackage{amssymb}
\usepackage{bm}
\usepackage{xcolor}

\begin{document}

\title{Thermofractals and the Nonextensive Finite Ideal Gas}

\author{A. Deppman$^{1}$} \email{deppman@if.usp.br}
\author{J. A. S. Lima$^{2}$} \email{jas.lima@iag.usp.br} 
\affiliation{$^{2}$Instituto de F\'isica, Universidade de S\~ao Paulo, São Paulo, Brasil}
\affiliation{$^{1}$Departamento de Astronomia (IAG-USP), Universidade de S\~ao Paulo, 05508-090 São Paulo SP,  Brasil}


\begin{abstract}
        {The underlying connection between the degrees of freedom of a system and its nonextensive thermodynamic behavior is addressed. The problem is handled by starting from a thermodynamical system with fractal structure and its analytical reduction to a finite ideal gas.  In the limit where the thermofractal has no internal structure, it is found that it reproduces the basic properties  of a nonextensive ideal gas with a finite number of particles as recently discussed (Lima \& Deppman, Phys. Rev. E 101, 040102(R) 2020). In particular, the entropic $q$-index is calculated in terms of the number of particles both for the nonrelativistic and relativistic cases. In light of such results, the possible nonadditivity or additivity of the entropic structures are also critically analysed  and new  expressions to the entropy (per particle) for a composed system of thermofractals and its limiting case are derived.} 
\end{abstract}
\maketitle

\noindent{\it 1. Introduction.} The last few decades have witnessed fast developments involving theoretical approaches to complex systems and fractality, including an increasing  understanding of the entropy concept, a state variable originally proposed by Clausius in the framework of classical thermodynamics, and, subsequently, extended by Boltzmann and Gibbs \cite{LBolt,Gibbs}. 

An extension to quantum systems was proposed  by von Neumann, while Shannon investigated its connection to the information theoretical domain \cite{VN32,S48}. Jaynes argued that if the von Newmann and Shanon entropy concept is accepted as a trustworthy measure of statistical uncertainty, the entropy becomes a primitive concept even more important than the energy itself \cite{J57}. 


A new  and deep investigation involving entropy, its physical meaning and implications for different fields, was initiated in 1988 through a seminal paper by Tsallis~\cite{Tsallis1988}. Assuming that entropy is a fundamental concept, Tsallis extended the traditional BG expression to  include processes and systems that could not be described by an additive entropy.  

In the homogeneous microcanonical ensemble, for instance, Tsallis entropy and its BG limit can be written as:  
\begin{equation}
S_q = k_B\,ln_q\,W\,\,\,(T)\,, \,\,\,\,\, \lim_{q\rightarrow 1}S_q=k_B\,ln W  \,\,\, (BG)\,,
\end{equation}
where $W$ is the number of microscopic accessible states and the q-index quantifies  how  Tsallis' entropy departs from the extensive BG statistics. In the above expression we have defined the q-logarithm function, 
\begin{equation}
 ln_q\, W = \frac{W^{q-1}-1}{q-1}\,,
\end{equation}
so that in the limit $q \rightarrow 1$, $ln_q\,W =ln W$. It is simple to proof that its inverse is the q-exponential function 
\begin{equation}
 e_q(W)=[1+(q-1)W]^\frac{1}{q-1}\,,
\end{equation}
which in the limit $q \rightarrow 1$, $e_q(W) =exp (W)$. The Tsallis q-entropy is nonadditive in the sense that
\begin{equation}
 S_q(A+B)= S(A) + S(B) + k_B^{-1}(1-q)S(A)S(B)
\end{equation}
implying the existence of correlations ($q>1$) or anti-correlations ($q<1$) in the system.

Motivated by the nonextensive statistics, many investigations have been performed on the concept of entropy \cite{BeckCohen,Thurner,Tempesta,Kalogeropoulos} or on the features that a system must present in order to follow the generalized statistics \cite{Wilk2007,Wilk2009,Biro,Deppman2016,Borland},  and it was shown that the Tsallis statistics can also be derived from considerations about the kinetic behavior describing gases and plasma \cite{Borland,Ademir1,LRS2000,Ademir2}. 

Recently, it has been argued that Tsallis statistics is not only more general than BG statistics, but also more fundamental, since the simplest thermodynamical system one can imagine, namely, the few body ideal gas, follows Tsallis statistics with $q$ depending only on the number, $N$ of components in the system, that is, $q=1+2/(3N)$ \cite{LimaDeppman}. In a diferent line of investigation,  it was also shown that the system called thermofractal can give rise to nonextensive thermodynamics effects even for large systems \cite{DeppmanMegiasPhysics, Deppman2016}. The algebra of the group of transformations of the thermofractal and the q-algebra are isomorphic \cite{DeppmanPhysics2021}.

A few different approaches have been proposed to show how the nonextensive behavior emerges from the properties of the system, as temperature fluctuation \cite{Wilk2007,Wilk2009}, finite size of the system~\cite{Biro,LimaDeppman} or fractal structure \cite{Deppman2016}. It has already been shown that finite systems~\cite{BiroGroup}, as well as thermofractals \cite{DFMM} are regulated by the same temperature fluctuations leading to a generic Tsallis statistics \cite{Wilk2009}. 

In this \textit{Rapid Communication}  we show that the nonextensive few-body ideal gas can be obtained as a limiting case of thermofractal systems. In particular, this may indicate  that all three approaches mentioned above, namely, temperature fluctuations, small system and thermofractals,  are indeed signatures of the the same feature responsible for the emergence of nonextensive effects. New nonadditive formulas for the total entropy involving  nonrelativistic and relativistic  subsystems with different number of particles are also derived and some basic consequences critically discussed.

\noindent\textit{2. Thermofractals.} The so-called thermofractals are systems in thermodynamical equilibrium presenting the following basic properties \cite{Deppman2016, FractalYMF}:
 
 [1.] The total energy is given by
 \begin{equation}
  U=K+E\,,
 \end{equation}
where $K$ corresponds to the kinetic energy of $\tilde{N}$ constituent subsystems and  $U$ describes the internal energy of those subsystems, which are endowed with an internal substructure.

[2.] The constituent particles are thermofractals which can be divided in two sublasses: Type I and Type II. 
 In the former case, the average ratio $\langle U \rangle /\langle K \rangle$ is constant for all the subsystems while in the later the constant ratio is $\langle U \rangle /\langle E \rangle$. For each subclass, the corresponding  associated ratio, $U/K$ and $U/E$,  can vary according to a distribution which is self-similar or self-affine, $P(U)$. This means that  at different levels of the subsystem hierarchy the distribution of the internal energy are equal (proportional) to those in the other levels.

[3.] At some level $n$ in the hierarchy of subsystems the phase space is so narrow that one can neglect their internal structure and assume the following expression to the probability:
\begin{equation}
 P(U_n) \, dU_n=\rho \, dU_n\,,
\end{equation}
\noindent with $\rho$ being independent of the internal energy $U_n$.

It can also be demonstrated that any system endowed with these three properties follows a Tsallis probability distribution determined by the q-exponential densities \cite{Deppman2016,DeppmanMegiasPhysics,FractalYMF} 
\begin{equation}
 P(u)=A \left[1 \pm (q-1)\frac{u}{\lambda}\right]^{\mp 1/(q-1)}\,, \label{power-law1-2}
\end{equation}
where $A$ is a normalization constant and the negative or positive signs in the exponent are associated to Type-I and Type-II, respectively, and  $u=U/\tilde N$.  The argument  $(q-1)u/\lambda=\chi$ in the distribution  (\ref {power-law1-2}) is a scale-free variable. Recall that for the Type-I thermofractals, $\chi=U/K$,  while for those of Type-II, $\chi=U/E$. 

In the formula above, the entropic index, $q$, is related to the properties of the fractal structure. It depends on the number of components of the thermofractal,  $\tilde{N}$,  and also of their internal structure obeying the relation \cite{Deppman2016}
\begin{equation}
q-1=\frac{1-\nu}{\alpha}\,, \label{qnu}
\end{equation}
where  $\nu$ is the fraction of thermofractal components contained in $\tilde{N}$ when their internal structure is not considered.

The distribution of energy among the components at each level of the thermofractal is exclusively determined as a function of scale-free variable, $\chi$, and  follows the q-exponential distribution. The main features of thermofractals have been discussed in the last few years \cite{Deppman2016,DFMM,Deppman_Universe}, and although those works were dedicated to the study of the type-I thermofractals the results can be straightforwardly extended to type-II. 

In what follows we will focus our attention on the parameter $\nu$ and its role in the thermofractal structure. To begin with we recall that the Tsallis q-exponential in Eq.(\ref{power-law1-2}) can be written in terms of the Euler Gamma Function integral form given by
\begin{equation}
 \begin{split}
P(u^{(n)},u^{(n-1)})  =  & A
\,
\left[1-\frac{u^{(n)}}{u^{(n-1)}} \right]^{\alpha} 
\left[\tilde P(u^{(n-1)})\right]^{\nu}  \,, \label{efdistribution}
 \end{split} 
\end{equation} 
where the parameter $\alpha=4\tilde{N}$ for relativistic systems and $\alpha=3\tilde N/2$  for a   non-relativistic system.

In the equation above, $u^{(n-1)}$ and $u^{(n)}$ are, respectively, the energy of thermofractals at the $(n-1)$th and $n$th levels of the hierarchic structure (for more details see Ref.~\cite{DFMM}). The term $\tilde P(u^{(n-1)})$ includes the energy fluctuation of the thermofractal at the level $n-1$. The $\tilde N$ thermofractals at the level $n$ are components of the thermofractal at the level $n-1$, and includes a fraction of the total number of degrees of freedom of that thermofractal.

The second property of thermofractals indicates that the energy distribution of these systems is self-similar, and scales with the energy. It means that the function describing the energy fluctuation of the thermofractal at the level $n-1$ must be identical to the that function describing the energy fluctuation of the thermofractal at the level $n$, as far they are expressed in terms of the scale invariant variable, $\chi$. This assertion is mathematically expressed as
\begin{equation}
 P\left(\frac{u^{(n)}}{u^{(n-1)}}\right)=\tilde P\left(\frac{u^{(n-1)}}{u^{(n-2)}} \right)\,. \label{eqn:selfsimilarity}
\end{equation}
Since the identity above is valid for any pair of consecutive levels in the thermofractal structure,  it is convenient to rewritten it  a level independent form through the following ratio
\begin{equation}
 \frac{u}{\Lambda}=\frac{u^{(n)}}{u^{(n-1)}}=\frac{u^{(n-1)}}{u^{(n-2)}} \,.
\end{equation}


\noindent\textit{3. Nonextensive ideal gas as a thermofractal limit.} An interesting limiting case of thermofractals happens for $\nu=0$, that is, when there are no components beyond the $\tilde{N}$ components already taken into account in the calculations. Such a description corresponds exactly to the case where the components do not present any internal structure, and, as such, one would expect a reduction of the thermofractal system to an ideal finite gas with $\tilde{N}$ particles. 

We observe that the thermofractal must satisfy~\eqref{efdistribution} and~\eqref{eqn:selfsimilarity} simultaneously, so we can write the basic functional equation

\begin{equation}
 P\left(\frac{u}{\Lambda}\right)=\left[1-\frac{u}{\Lambda} \right]^{\alpha} 
\left[P\left(\frac{u}{\Lambda}\right)\right]^{\nu} \,. \label{eqn:recurrent}
\end{equation}
which only possible solution reads:
\begin{equation}
 P(u)=A \left[1 \pm (q-1)\frac{u}{\lambda}\right]^{\mp 1/(q-1)}\,, \label{eqn:Tsallis}
\end{equation}
with $q$ being determined by Equation~\ref{qnu}.


Inserting Eq. (\ref{qnu}) with $\nu=0$ into Eq. (\ref{power-law1-2}) we find: 
\begin{equation}
 \left[1\pm (q-1)\frac{u}{\lambda}\right]^{\mp 1/(q-1)}=  \left(1 \pm \frac{u}{\Lambda}\right)^{\alpha}   \,, \label{DoubleTsallisDist}
\end{equation}
where $\Lambda=\alpha\,\lambda$ for the relativistic and non-relativistic gases. If we further restrict the calculations to the case of non-relativistic type-II thermofractals, the probability distribution results in
\begin{equation}
 P(U)= A \left(1 - \frac{u}{\Lambda}\right)^{ 3 \tilde N/2}\,,
\end{equation}
which is exactly the probability distribution obtained in Ref.~\cite{LimaDeppman}, where an analysis of the finite ideal gas was recently performed based  on the Liouville theorem.

Thermofractals may also be reduced to a finite ideal gas  by a condition less restrictive than imposing $\nu=0$. In fact, the fundamental aspect allowing the  thermofractal reduction to an ideal gas is that the components have no internal structure. In this case,  $\nu \ne 0$ means that there are missing degrees of freedom in the system but, in this case, they cannot be internal degrees of freedom of any component already considered. Hence, these hidden degrees of freedom come from particles in the system which are not included in the basic calculations.  

The discussion made so far shows that we can have non-extensive systems characterized by different entropic index. It is intersting, therefore, to observe what happens with the entropy of a system that is composed by a mixture of systems with different values for $q$. Using Eqs Equations~\ref{qnu} and~\ref{eqn:Tsallis}, Eq.~\ref{eqn:recurrent} can be written as
\begin{equation}
 \begin{split}
P(u,u'')  = & A' \left[1-(q'-1)\frac{u'}{\lambda'}\right]^{1/(q'-1)} \times \\ & A''
\left[1-(q''-1)\frac{u''}{\lambda''}\right]^{1/(q''-1)}  \label{efdistribution3}
 \end{split}
\end{equation}
where we have introduced the quantities $q'$, $q''$, $\lambda'$ and $\lambda''$ such that
\begin{equation}
 \begin{cases}
  & q'-1=\frac{1}{\alpha'} \,;\, \lambda'=(q'-1)\Lambda \\
  & q''-1=\frac{1}{\alpha''} \,;\, \lambda''=(q''-1)\Lambda
 \end{cases}
\end{equation}
and $A'A''=A$. In this case we have
\begin{equation}
 \begin{split}
 A \left[1-(q-1)\frac{u}{\lambda}\right]^{1/(q-1)}= & A' \left[1-(q'-1)\frac{u'}{\lambda'}\right]^{1/(q'-1)} \times \\ & A''
\left[1-(q''-1)\frac{u''}{\lambda''}\right]^{1/(q''-1)} \,. \label{NormalizationRelation}
 \end{split}
\end{equation}
The equation above represents the probability independence of the systems. As we will see below, this independence does not prevent the non-additivity of the entropy, a characteristic of the Tsallis statistics

To completely prove the validity of Eq.~(\ref{NormalizationRelation}) in the context of thermofractals,  we need to show that the normalization constants $A$, $A'$ and $A''$ follow the same function of $q$. Consider that a system of $N$ particles contained in a volume $V$ is  partitioned into two subsystems with $\tilde{N}$ and $N'$, and let $M(N,U)\, dU$ be the number of configurations, that can be mathematically expressed as
\begin{equation}
 M(N,u)\, du=\rho V \int_{-\infty}^{\infty} d^{3N}p \int_{-\infty}^{\infty} d^{3N}x \delta(U-u)  \,, \label{initialformula}
\end{equation}
where $\rho(x,p)=\rho$ is the density of states per unit of phase-space volume, assumed to be constant. 
The total energy if $U=p^2/(2m)$, where $p^2=\sum {\bf p}_i^2$ and ${\bf x}_i$  are the postion if the particles in the coordinate space.

In Eq.~(\ref{initialformula}) it is clear that we calculate the volume of the phase-space for $N$ particles, since the topological dimension of the space is $3N$, and the Dirac's delta function restrict the available gas configurations to those with total energy $U$. Assuming that the system is free of external forces and particles can move freely in a volume $V$, Eq.~(\ref{initialformula}) can be integrated in the coordinate variables resulting in a constant fact $V^N$. The integration over the momentum variables can be performed considering that $p$ is the radius of a hyper-sphere in the $3N$ dimensional momentum space, and using the transformation $d^{3N}p \rightarrow dp \, p^{3N-1} S(3N)$ where
\begin{equation}
 S(n)=\frac{2 \pi^{n/2}}{\Gamma(n/2)}\,, \label{Sterm}
\end{equation}
with $\Gamma(x)$ being the Euler's Gamma Function. 

Using the transformation above in Eq.~(\ref{initialformula}) we obtain
\begin{equation}
 M(N,u)\, du=\rho V^N\int_0^{\infty} dp \, p^{3N-1} S(3N) \, \delta(U-u) \, du\,, \label{sphericalinitialformula}
\end{equation}
where the dependence of the delta function on $p$ is given by the energy $u$. We write $p$ in terms of $U$ and integrate to obtain
\begin{equation}
 M(N,u)\, du=\rho V^N \frac{(2\pi m)^{3N/2}}{\Gamma(3N/2)}u^{3N/2-1} \, du \,, \label{NConfigs}
\end{equation}
where Eq.~(\ref{Sterm}) was used.  Since we are assuming no internal structure for the components, the total energy is completely determined by the degrees of freedom considered in our calculations and the infinitesimal variation, $dE$, can be suppressed. The quantity $M(N,u)$, then, indicates the number of configurations.

If we introduce further constraints to the system, and consider the number of configurations for the system with $N$ particles and total energy $u$ where a part  of the total energy between $\varepsilon$ and $\varepsilon+d\varepsilon$ is shared by a number $\nu$ of the particles in the system, $M_ {\nu}^{\varepsilon}(N,u)$. The corresponding number density is given by
\begin{equation}
 \begin{split}
 M_ {\nu}^{\varepsilon}(N,u)  =\rho \, V^N \int_0^{\infty} d^{3\nu}p \int_0^{\infty}d^{3(N-\nu)}p \delta \left(u'-(u-\varepsilon)\right)\,, \label{referenceformula}
 \end{split}
\end{equation}
where
\begin{equation}
 u'=\sum_{i=\nu+1}^{N}\frac{{\bf p}_i^2}{2m}
\end{equation}
is the energy available to the remaining system with $N-\nu$ particles. Using the substitution of variables from momentum to energy we obtain
\begin{equation}
 \begin{split}
 & M_ {\nu}^{\varepsilon}(N,u)=  \rho \left[V^{\nu} \frac{(2\pi m)^{3\nu/2}}{\Gamma(3\nu/2)} \int_0^{\infty}\varepsilon^{3\nu/2-1} d\varepsilon\right] \times \\ &  \left[V^{(N-\nu)} \frac{(2\pi m)^{3(N-\nu)/2-1}}{\Gamma(3(N-\nu)/2)} \int_0^{\infty} u'^{3(N-\nu)/2-1} du'\right] \times \\ &\delta \left(u'-(u-\varepsilon)\right)\,, \label{clusterformula}
 \end{split}
\end{equation}
what shows that
\begin{equation}
 M_ {\nu}^{\varepsilon}(N,u)=M(\nu,\varepsilon)\times M(N-\nu,u')\,, \label{additivity}
\end{equation}
with the condition that $u'+\varepsilon=u$. Observe the similarities between the equation above and Eq.~(\ref{NormalizationRelation}).

Summming Eq.~(\ref{clusterformula}) for $\nu=1$ to $N$ we have, after rearranging appropriately all factors,
\begin{equation}
 \begin{split}
 & \sum_{\nu=1}^{N}M_ {\nu}^{\varepsilon}(N,u)= \rho V^N\frac{(2\pi m)^{3N/2}}{\Gamma(3N/2)} \times \\ & \sum_{\nu=1}^{N} \frac{\Gamma(3N/2)}{\Gamma(3\nu/2)\Gamma(3N/2-3\nu/2)} \varepsilon^{3\nu/2-1}u'^{3(N-\nu)/2-1} \,. \label{totalnormalization}
  \end{split}
\end{equation}
But
\begin{equation}
 \begin{split}
 & \sum_{\nu=1}^{N} \frac{\Gamma(3N/2)}{\Gamma(3\nu/2)\Gamma(3N/2-3\nu/2)} \varepsilon^{3\nu/2}u'^{3(N-\nu)/2}= \\ &(\varepsilon+u')^{3N/2-2}\,,
  \end{split}
\end{equation}
therefore
\begin{equation}
 \sum_{\nu=1}^{N}M_ {\nu}^{\varepsilon}(N,u)=\rho \,V^N\frac{(2\pi m)^{3N/2}}{\Gamma(3N/2)}u^{3N/2-2}\,. \label{totalsum}
\end{equation}
The right hand side in the expression above is exactly the number of configuration available to the N particle gas with energy between $U$ and $U+dU$, according to Eq.~(\ref{NConfigs}), so we get the expected result
\begin{equation}
 M(N,u)=\sum_{\nu=1}^{N} M_{\nu}^{\epsilon}(N,u)\,,
\end{equation}
that inform us that when we consider configurations where an energy $\varepsilon$ is distributed among $\nu$ particles in a N-particle system with total energy E, then when we sum over all possible values for $\nu$ we get the total number of configurations available to the whole system.

In statistical physics it is more appropriate to use probability distributions instead of number of states. For the present case it is simple to go from the number density, $M(N,u)$ to probability density, $P(N,u)$ if one assumes that all states can be occupied with the same probability, and here this assumption will be used. The probability for a particular configuration with $\nu$ particles carrying the energy $\varepsilon$, its probability density can be obtained by dividing Eq.~(\ref{clusterformula}) by the total number of configurations for the system, $M(N,u)$, that is
\begin{equation}
 P(\nu,\varepsilon)=\frac{M_{\nu}^{\varepsilon}(N,u)}{M(N,u)}\,,
\end{equation}
resulting in
\begin{equation}
 \begin{split}
   P(\nu,\varepsilon) d\varepsilon= & \frac{\Gamma(3N/2)}{\Gamma(3\nu/2)\Gamma(3N/2-3\nu/2)} u^{-(3N/2-1)} \, \times \\  & \varepsilon^{3\nu/2-1}u'^{3(N-\nu)/2-1}  d\varepsilon \, du' \delta \left(u'-(u-\varepsilon)\right)\,.
 \end{split}
\end{equation}
Notice that with the normalization used here we attribute probability 1 to the system with energy between $u$ and $u+du$.
Integrating the equation above on $u'$ we get
\begin{equation}
 \begin{split}
 P(\nu,\varepsilon) d\varepsilon= & \frac{\Gamma(3N/2)}{\Gamma(3\nu/2)\Gamma(3N/2-3\nu/2)} \left(\frac{\varepsilon}{u}\right)^{3\nu/2-1} \times \\ &\left[1-\frac{\varepsilon}{u}\right]^{3(N-\nu)/2-1} \frac{d\varepsilon}{u}\,. \label{clusterdensity}
 \end{split}
\end{equation}
The expression above gives the probability to obtain, in an $N$ particles ideal gas with total energy $u$, a subset of $\nu$ particles with energy $\varepsilon$.

Comparing Eqs~(\ref{additivity}),~(\ref{totalnormalization}) and~(\ref{clusterdensity}), we observe that the normalization constants are consistently defined by
\begin{equation}
 \begin{cases}
  & A= V^{-N}(2\pi m)^{-3N/2}\Gamma(3N/2) \\
  & A'=  V^{-\tilde{N}}(2\pi m)^{-3N'/2}\Gamma(3\tilde{N}/2)\\
  & A''= V^{-N'}(2\pi m)^{-3N''/2}\Gamma(3N'/2)
 \end{cases}
\end{equation}
A thorough discussion about the special case where $\nu=1$ was presented in Ref.~\cite{LimaDeppman}. 

As far as the molecules of the gas occupy the same volume $V$ and have the same mass $m$, the terms that contain $V$ and $m$ can be removed from the normalization constants without loss of generality. Therefore, from Eq.~(\ref{NormalizationRelation}) and using the normalization constants determined above, we conclude that the normalized probability distribution is given by
\begin{equation}
 P(q,\lambda;\varepsilon)=\frac{1}{\Gamma\left(1/(q-1)\right)}[1-(q-1)(\varepsilon/\lambda)]^{-1/(q-1)}\,.
\end{equation}

At this point, it is interesting to analyze one additional case, that of $\nu=0$. As discussed above, this case corresponds to a gas where the $\tilde{N}$ particles represent a null fraction of the total system. This means that the system has an infinite number of components, according to Eq.~(\ref{qnu}). In this case one expect that the system behaves according to BG statistics. Indeed, for $\nu=0$ we obtain that $q=1$, thus indicating the limit when Tsallis statistics is identical to BG statistics.


\noindent\textit{Non-additivity of entropy.} Let us now analyse the parameter $\lambda$ which is considered here as a reduced scale. We first observe that the conditions $U=U'+U''$ and $\alpha=\alpha'+\alpha''$ must simultaneously be satisfied by the thermofractals also in the ideal gas limit. Note also that $\lambda \propto E/N$ for all partitions of the gas so that one may conclude that
\begin{equation}
 \frac{u}{\tilde N}=\frac{u'+u''}{\tilde N'+\tilde N''}\,,
\end{equation}
and using $\lambda'=u'/\alpha'$ and $\lambda''=u''/\alpha''$ we obtain that $ \alpha\lambda= \alpha'\lambda'+\alpha''\lambda''$, and in terms of the parameter $q$ we have
\begin{equation}
 \frac{\lambda}{q-1}=\frac{\lambda'}{q'-1}+\frac{\lambda''}{q''-1}\,. \label{lambdarelation}
\end{equation}
The above result shows that in this case, the equilibrium temperature (which plays the role of a scale variable in the system),  is uniquely defined and, as such,  can be interpreted as a partition of one single system. 

{\it At this point  one may ask: is the entropy of the full system  additive or nonadditive?} In order to answer that let us indicate by $p_q$ the probability $P(q,\lambda;\varepsilon)$ for some specific value of $\varepsilon$. It thus follows that Eq.~(\ref{additivity}) can be rewritten as
\begin{equation}
 p_q=p_{q'}p_{q''}\,.
\end{equation}
The relation above shows that the probability distributions of each partition is independent of the other. If $\sigma=\varepsilon/\lambda$, it follows that
\begin{equation}
 \sigma_q=\frac{1-p_q^{q-1}}{q-1}\,,
\end{equation}
which is, apart from a constant, the Tsallis entropy, $S_q$. The results above show that if $p_{q'}$ and $p_{q''}$ follow Tsallis distributions, the combined system is still non additive, in general. In fact, if we have the entropies $\sigma_{q'}$ and $\sigma_{q''}$ associated, respectivelly, to the probabilities, $p_{q'}$ and $p_{q''}$, is follows from the equation above that
\begin{equation}
 \sigma=\frac{q'-1}{q-1}\sigma'+\frac{q''-1}{q-1}\sigma''-\frac{(q'-1)(q''-1)}{q-1}\sigma'\sigma''\,. \label{generalnonadditivity}
\end{equation}
This is a straightfoward generalization of the well-known Tsallis' nonadditive entropic rule for a combination of systems with different entropic indexes, $q'$ and $q''$. 
In terms of the number of particles, we find
\begin{equation}
 \frac{\sigma}{\tilde \alpha}=\frac{\sigma'}{\tilde \alpha'}+\frac{\sigma''}{\tilde \alpha''}-\frac{\sigma'}{\tilde \alpha'}\frac{\sigma''}{\tilde \alpha''}\,, \label{generalnonadditivity}
\end{equation}
 an expression of general validity in the present context. As one may check, for a nonrelativistic finite gas, for instante, $\alpha = 2/3\tilde N$ ($\alpha' = 2/3 N'$, $\alpha'' = 2/3N''$), the above expression reduces to:

\begin{equation}
 \frac{\sigma}{\tilde N}=\frac{\sigma'}{\tilde N'}+\frac{\sigma''}{\tilde N''} -\frac{2}{3} \frac{\sigma'}{\tilde N'}\frac{\sigma''}{\tilde N''}\,, \label{generalnonadditivity}
\end{equation}
while for the relativistic case, $\alpha = 4{\tilde N}$ and so on, the only difference is that the numerical factor multiplying the last term above changes from 2/3 to 1/4. 

It should also be stressed that the proportionality relating the $\alpha$ parameters with the number of components in each system is the signature of the entropic nonadditivity with the particle number of the global system. 
In addition, for $N'= N''$, that is, when the merged systems have the same number of particles, and, as such, the same value for the parameter $q$, the total entropy still remains nonadditive (this is the particular case usually considered in the literature). As should be expected, the additive BG limit is obtained only when the standard thermodynamic limit is attained, that is, for $N'$ ($N''$) or both  approaching the infinite.



\noindent\textit{Discussion and conclusion.} In summary, we have demonstrated  that the finite ideal gas can be obtained as a limiting case of the so-called thermofractal, that is, a system endowed with fractal internal structure affecting its overall thermodynamical behavior. In such a limit, the thermofractal is formed by components with no internal structure ($\nu=0$) with the generic thermofractal entropic index  reducing to $q= 1 + 2/(3N)$, where $N$ is the number of point-like components of the system [cf. (\ref{qnu})].  It shouldbe stressed that such a q-index for an ideal finite gas was also recently derived from first principles based on the Liouville N-particle distribution function \cite{LimaDeppman}. Conversely,  this result also reinforces the internal consistency of the thermofractal concept. In principle, one would expect that the above result should naturally be recovered from any compelling thermofractal description, as long as its internal structure is neglected. 

In the finite ideal gas, the nonextensive behavior fade away with the growth  of the number of components ($N \rightarrow \infty$\,, $q \rightarrow 1$). In other words, the Tsallis nonextensive behavior happens only for small systems. However, the more general thermofractal structure seeds the emergence of nonextensivity in large systems. The present work also put in evidence an interesting  feature for generic thermofractal large system: although not unique, a limited number of degrees of freedom can be considered a basic mechanism  for the presence of nonextensivity.

Considering the present results under the light of other achievements in the literature about Tsallis distribution, we observe that finite ideal gas presents a distribution of the inverse temperature, $\beta=1/T$, that corresponds to a Euler-Gamma function. As it is known, systems where the temperature follows such distribution are described by Tsallis statistics. At light of these results, we also conjecture its relevance for some physical systems which described by a limited number of degrees of freedom. One has to keep in mind, however, that other sources of nonextensivity may exist and play a relevant role in the statistical properties of physical systems~\cite{Cirto}.

In conclusion, the few-body ideal gas can be obtained as a limiting case of the thermofractal of Type-II (see the basic properties in section 2). The result is in agreement with previous approaches, and  yields an entropic $q$-index  as a function of the number of particles in the ideal gas. We studied the composition law of two systems, and established a general law for the equilibrium temperature in the case of ideal gas with finite number of particles. The calculations also allow us to establish a general rule for relating the entropic $q$-index for composed systems.  We have analized our results and compared with previous ones derived in the literature, what enable us to discuss some deep aspects of Tsallis statistics. 
\vskip 0.5cm
\noindent {\bf Acknowledgements:}  A.D. is partially
supported by CNPq (304244/2018-0), INCT-FNA
(464898/2014-5), and FAPESP (2016/17612-7). J.A.S.L. is partially supported by
CNPq (310038/2019-7), CAPES (88881.068485/2014), and
FAPESP (LLAMA Project No. 11/51676-9).

\bibliographystyle{unsrt}

\end{document}